\documentclass[aip,graphicx, amsmath,amssymb,reprint]{revtex4-1}
\usepackage{graphicx}
\draft 
\usepackage{mhchem}
\usepackage{filecontents}
\usepackage{subcaption}
\usepackage{comment}
\usepackage{svg}
\usepackage{cleveref}

\begin{document}


\title{Enhancing All-Optical Switching of Magnetization by He Ion Irradiation} 



\author{Pingzhi Li$^*$}
\email[]{p.li1@tue.nl}
\affiliation{Department of Applied Physics, Eindhoven University of Technology, P. O. Box 513, 5600 MB Eindhoven, The Netherlands}
\author{Johannes W. van der Jagt}
\affiliation{Spin-Ion Technologies, 10 boulevard Thomas Gobert, 91120 Palaiseau, France}
\affiliation{Universit\'e Paris-Saclay, 3 rue Juliot Curie, 91190 Gif-sur-Yvette, France}
\author{Maarten Beens}
\affiliation{Department of Applied Physics, Eindhoven University of Technology, P. O. Box 513, 5600 MB Eindhoven, The Netherlands}
\author{Julian Hintermayr}
\affiliation{Department of Applied Physics, Eindhoven University of Technology, P. O. Box 513, 5600 MB Eindhoven, The Netherlands}
\author{Marcel Verheijen}
\affiliation{Department of Applied Physics, Eindhoven University of Technology, P. O. Box 513, 5600 MB Eindhoven, The Netherlands}
\author{Ren\'e Bruikman}
\affiliation{Department of Applied Physics, Eindhoven University of Technology, P. O. Box 513, 5600 MB Eindhoven, The Netherlands}
\author{Beatriz Barcones}
\affiliation{NanoLab@TU/e, Eindhoven University of Technology, P. O. Box 513, 5600 MB Eindhoven, The Netherlands}
\author{Rom\'eo Juge}
\affiliation{Spin-Ion Technologies, 10 boulevard Thomas Gobert, 91120 Palaiseau, France}
\author{Reinoud Lavrijsen}
\affiliation{Department of Applied Physics, Eindhoven University of Technology, P. O. Box 513, 5600 MB Eindhoven, The Netherlands}
\author{Dafin\'e Ravelosona}
\affiliation{Spin-Ion Technologies, 10 boulevard Thomas Gobert, 91120 Palaiseau, France}
\affiliation{Centre de Nanosciences et de Nanotechnologies, CNRS, Universit\'e Paris-Saclay, 10 boulevard Thomas Gobert, 91120 Palaiseau, France}
\author{Bert Koopmans}
\affiliation{Department of Applied Physics, Eindhoven University of Technology, P. O. Box 513, 5600 MB Eindhoven, The Netherlands}


\date{\today}

\begin{abstract}

All-optical switching (AOS) of magnetization by a single femtosecond laser pulse in Co/Gd based synthetic ferrimagnets is the fastest magnetization switching process. On the other hand, He ion irradiation has become a promising tool for interface engineering of spintronic material platforms, giving rise to significant modification of magnetic properties. In this paper, we explore the use of He ion irradiation to enhance single pulse AOS of Co/Gd bilayer-based synthetic ferrimagnets. The intermixing of the constituent magnetic layers by He ion irradiation was both numerically simulated and experimentally verified. We theoretically modelled the effects of intermixing on AOS using the layered microscopic 3-temperature model and found that AOS is enhanced significantly by breaking the pristine Co/Gd interface through intermixing. Following this notion, we studied the threshold fluence of AOS as a function of He ion irradiation fluence. We found that the AOS threshold fluence can be reduced by almost 30\%. Our study reveals the control of AOS by He ion irradiation, which opens up an industrially compatible approach for local AOS engineering. 
  
\end{abstract}


\maketitle 


\begin{figure*}
\centering
\includegraphics[scale=1.1]{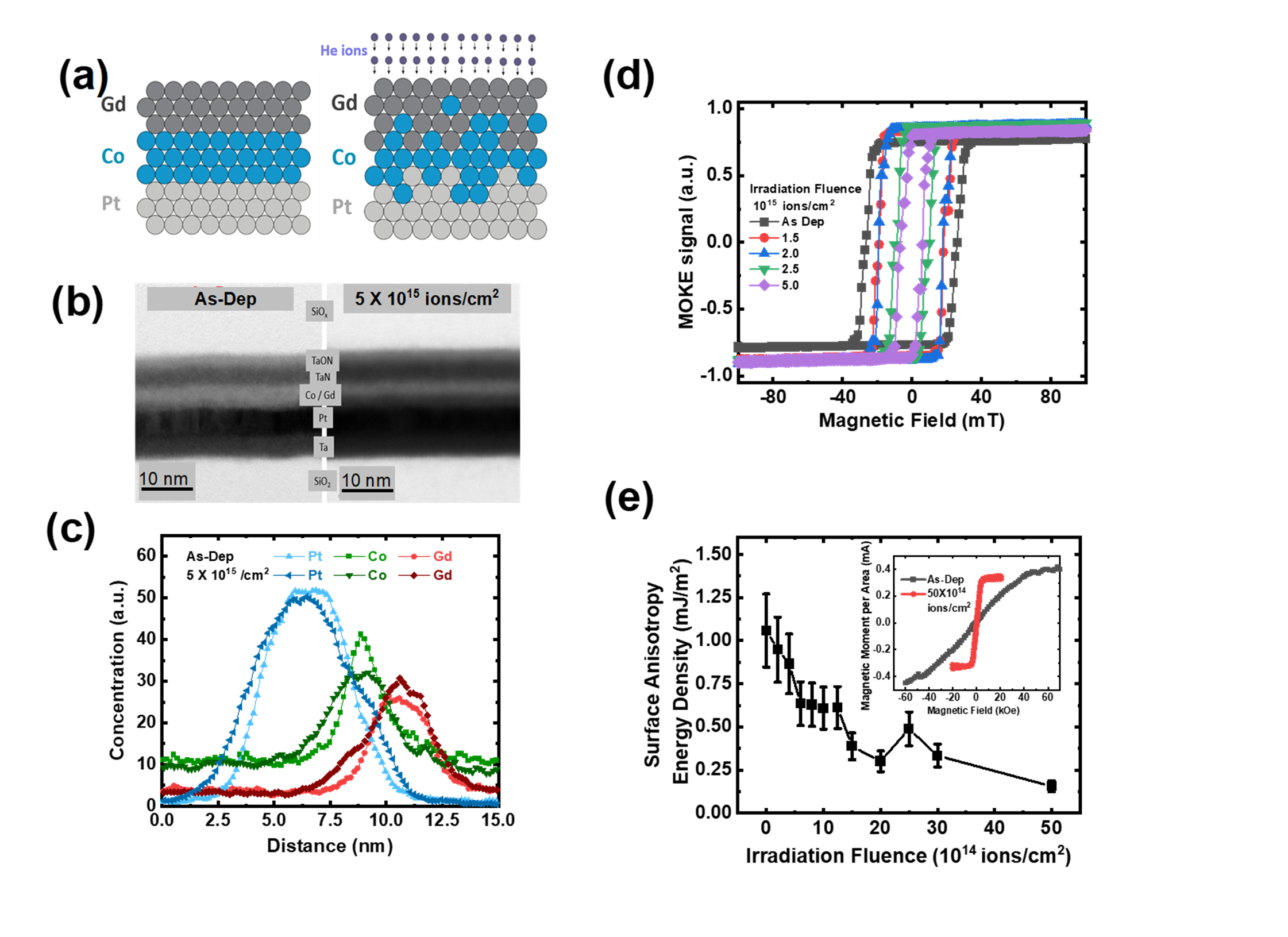}
\caption{(a) The schematic illustration of the intermixing effect induced by the He ion irradiation. 
    (b) Cross-section TEM image of the as-deposited sample (As-Dep) and the sample treated with an irradiation fluence of 50 $\times$ 10$^{14}$ ions/cm$^2$. The scale bar as well as the label for each layer is marked in the figure.  
    (c) Depth profile of the element mapping (Pt, Co, Gd) from EDX measurements for both As-Dep and irradiated sample with fluence of 5$\times$10$^{15}$ ions/cm$^2$. 
    (d)  Hysteresis loops, characterized by the polar-MOKE at the same field scanning speed, of As-Dep and irradiated sample with various irradiation fluence. 
    (e) The surface anistropy energy density characterized from the VSM-SQUID measurement in the hard axis direction. The magnetic moment as a function of the field for As-Dep and the sample with irradiation fluence of 5$\times$10$^{15}$ ions/cm$^2$ is shown in the inset.
     }
\label{fig:statics}
\end{figure*}

It is well-known that single pulse all-optical switching (AOS) of the magnetization in 3d-4f ferrimagnets is the fastest and one of the least dissipative ways of magnetization switching\cite{Kimel_AOS_Review2019,Polley:2022ws,Radu:2011aa,Ostler:2012aa,El-Ghazaly:2020aa}. This mechanism not only offers means for ultrafast memory operations, but also gives rise to integration between integrated photonics and spintronics\cite{Lalieu:2019aa, Polley:2022ws}, in which the 3d-4f ferrimagnets should prevail \cite{Kim:2022aa}.
Initially observed in a 3d-4f alloy systems \cite{Ostler:2012aa}, recently, it was found that the Co/Gd bilayer\cite{Lalieu:2017aa,Li:2022aa} based synthetic ferrimagnets exhibit likewise this ultrafast (ps switching time scale \cite{Peeters:2022uo, Luding:2022ur}) single pulse AOS process. The switching threshold fluence for this system is low\cite{Lalieu:2017aa,Li:2021wr,Hees:2020aa}. On the other hand, the high process tolerance\cite{Wang:2020ab} of this system also gives rise to an AOS-switchable magnetic tunnel junction with a high tunnel magneto-resistance \cite{Luding:2022ur} as the only material platform with such successful demonstration up today.

Particularly, a unique merit of this Co/Gd material platform is that the presence of AOS is not composition dependent \cite{Beens:2019aa}, unlike GdFeCo alloys \cite{Ostler:2012aa} and [Co/Tb]$_n$ multilayer systems\cite{Aviles-Felix:2020aa}, where only a narrow composition window of $\pm 2 \%$ from magnetic compensation is allowed \cite{Beens:2019aa} for AOS. Such a difference is rooted in the AOS reversal mechanism \cite{Gerlach:2017aa,Beens:2019aa}. Conventionally, for alloy systems, the switching relies on the angular momentum exchange \cite{Mentink:2012aa, Beens:2019aa} between 3d ferromagnet (Co) and Gd when the Co magnetic moment is quenched to zero while a proper amount of Gd is not fully demagnetized yet\cite{Radu:2011aa}. This results in the switching of Co magnetization, which is followed by the switching of Gd via the antiferromagnetic exchange coupling between Co and Gd. 
These properties of an alloy impose significant challenges on the possible wafer scale integration and leave no room for tunability of its AOS properties. On the other hand, for the Co/Gd layered system, the strict ratio between Co and Gd is lifted by the spatial asymmetry\cite{Beens:2019aa,Gerlach:2017aa}, as the switching takes place first at the Co/Gd interface which then initiates the switching of the rest of the ferromagnet. Following this notion, it was found that the AOS in Co/Gd based synthetic ferrimagnetic systems is independent of composition \cite{Li:2022aa,Beens:2019aa,Lalieu:2017aa,Li:2021wr} making it one of its unique properties. This creates an additional degree of freedom for nanoscale engineering of the properties of AOS.   

One of the methods to utilize this degree of freedom is to explore the intermixing at the Co and Gd interface \cite{Beens:2019ab,Wang:2020ab}. 
It was found that the threshold fluence of AOS can be largely reduced when the sharp interface between Co and Gd is replaced by a very thin layer of CoGd alloy. Interestingly, AOS still persists, despite the fact that the composition of this intermediate alloy is not the value pertaining AOS for the stand-alone alloy. Compared to the case with a pure interface, the switching mechanism remains the same, however, the initial switching at the interface region requires less energy.  Such a property can be experimentally verified as well as adopted for application purposes. 

To this end, it has been discovered that controlled promotion of intermixing in spintronic material systems can be realized by He ion irradiation\cite{Fassbender:2004vf}. Furthermore, local control can be realized by irradiation through a mask\cite{Chappert:1998to,Devolder:2001vw}. Such a post-growth process was shown to allow for a dramatic degree of modification of interfacial spintronic effects\cite{Herrera-Diez:2020aa,Zhao:2019aa, Devolder:2013aa, Zhao:2020wd, Krupinski:2021,Juge:2021vd}, including the perpendicular magnetic anisotropy (PMA), Dzyaloshinskii-Moriya interaction, and damping, which are critical parameters for various types of spintronic applications. This process can be potentially up-scaled to standard wafer-scale processing.


So far, He ion irradiation has not yet been adopted in the field of single pulse AOS, despite being highly explored already in the field of spintronics and optically driven domain wall motion \cite{El-Hadri:2018aa}. In this work, we study the effect of He ion irradiation on AOS of a Co/Gd bilayer-based synthetic ferrimagnet. We first present the physical evidence of the layer intermixing as a result of He ion irradiation as well as its the magneto-static properties. We then discuss our theoretical microscopic simulation for the intermixing effects in Co/Gd on AOS based on the numerically modeled intermixing profile, where we found AOS switching energy decreases with increased extend of intermixing. We further experimentally show that, as the irradiation fluence increases, the threshold fluence of AOS reduces by almost 30 $\%$. 
Our study bridges the link between the single-shot AOS and He ion irradiation, which paves the way for local and wafer-scale control of AOS properties, potentially useful for AOS threshold energy engineering.

To conduct the study as mentioned above, we deposited Ta(4)/Pt(4)/Co(1)/Gd(3)/TaN(4) (from bottom to top, thickness in parenthesis in nm)  on top of a Si/SiO$_2$(100) substrate using DC magnetron sputtering at a base pressure of 10$^{-9}$ mbar. The magneto-static properties of this material system have already been described in Ref. \cite{Lalieu:2017aa,Wang:2020aa}. The sample is further processed using a Helium-S system from Spin-Ion Technologies with uniform He$^{+}$ ion irradiation with an energy of 15 keV, and irradiation fluences up to 5 $\times$ 10$^{15}$ ions/cm$^{2}$.

\begin{figure*}[t]
\centering
\includegraphics[scale=0.6]{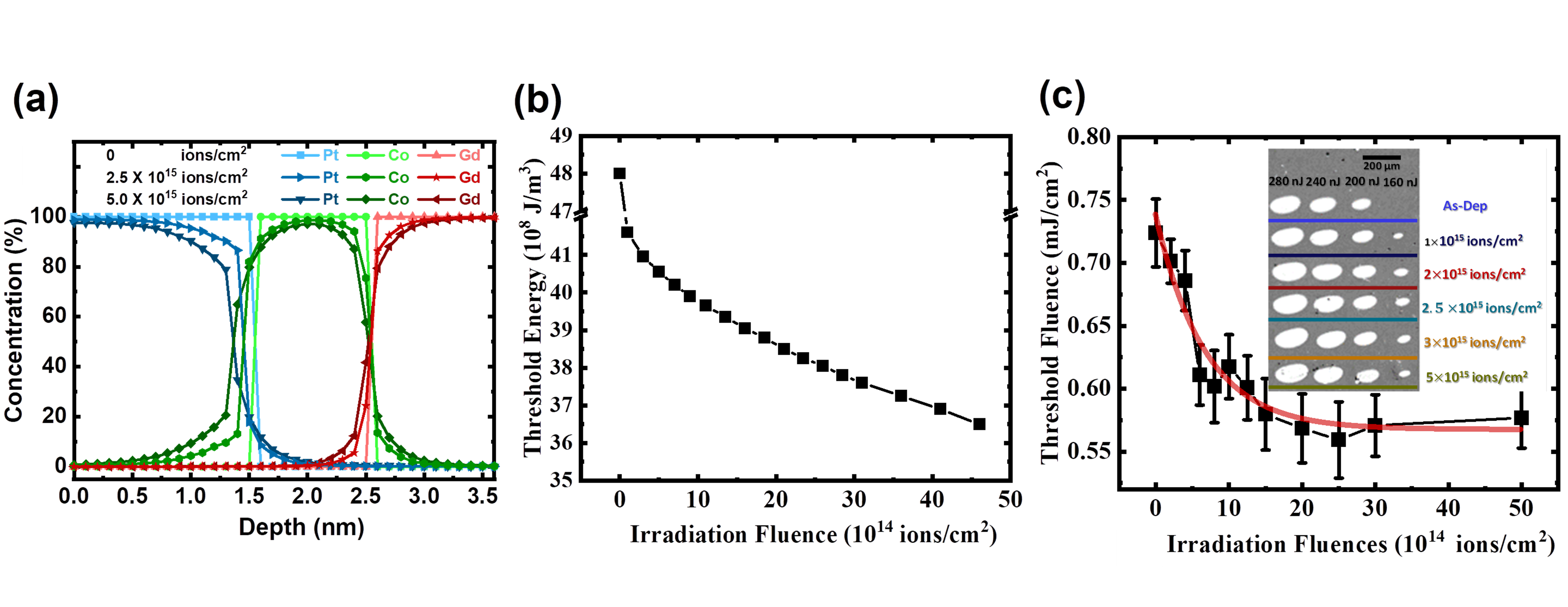}
\caption{(a) The depth profiles of the concentration of Pt, Co and Gd, for various irradiation fluences, which are obtained from TRIDYN simulations. 
    (b) Simulated threshold AOS energy of Co/Gd obtained from M3TM as a function of irradiation fluence, the concentration profiles obtained from TRIDYN simulations.  
    (c) AOS threshold fluence of Ta(4)/Pt(4)/Co(1)/Gd(3)/TaN(4) with different He$^{+}$ irradiation fluences. The guide of the eye is plotted as a red curve. Inset shows some example Kerr images of domains for samples with different irradiation fluences, created by a single linearly polarized laser pulse of different pulse energies.  
     }
\label{fig:AOS}
\end{figure*}

The irradiation is known to induce intermixing\cite{Fassbender:2004vf}, which is schematically illustrated in Fig. \ref{fig:statics}a. The accelerated He ions release their kinetic energy upon colliding with the  metal atoms in the film, which have much higher charge numbers and mass. This action agitates the atoms of the metallic layers such that the (positive) enthalpy of intermixing is released, promoting an intermixing between metal layers. By choosing a kinetic energy of 15\,keV\cite{Rettner:2002wr}, the majority of the He ions will be implanted into the substrate instead of remaining in the film, while still an appreciable amount of intermixing is induced at the metallic interfaces.
A physical evidence of induced intermixing can be visualized in the bright field transmission electron microscope (TEM) image shown in Fig. \ref{fig:statics}b, where the cross section of an as-deposited sample is compared to one irradiated by an exposure irradiation fluence of $5\times10^{15}$\,ions/cm$^2$. Pt/Co and Co/Gd interfaces in the as-deposited stack are well separated, as found in comparable transition--metal/Gd interfaces\cite{Hintermayr:2021,Hintermayr:2019}. A darker region on top of Pt and Gd can be observed in the irradiated sample as a result of intermixing at Pt/Co, Co/Gd, and Gd/TaN interfaces. Further evidence can be found in the elemental mapping of the energy dispersive X-Ray analysis (EDX) of the samples as deposited and irradiated with a fluence of 5$\times$10$^{15}$ ions/cm$^2$ as shown in Fig. \ref{fig:statics}c (concentration of He is below the resolution limit). The elements in the irradiated sample clearly have a broadened elemental distribution with respect to the thickness, which qualitatively indicates intermixing. Here, we note that the results shown in Fig. \ref{fig:statics}c allow for qualitative comparison only and do not reflect a direct measure of the real element concentration, due to the finite spot size of the electron beam and signal from the background, such as from the detector. 

We further investigated the magneto-static properties of the samples as a function of He ion irradiation fluence using the magneto-optic Kerr effect (MOKE). We first measured the hysteresis loop obtained using a polar MOKE configuration, some examples of which are shown in Fig. \ref{fig:statics}d. Here we observed square hysteresis loops both for as deposited and irradiated samples with a fluence up to 5$\times$10$^{15}$ ions/cm$^2$, which suggests all samples exhibit PMA. It is found that the coercive field is reduced as the He ion irradiation fluence increases. This is associated with the reduction of PMA induced at the Pt/Co interface due to increased intermixing\cite{Devolder:2000ub}. 
We further quantified the interfacial anisotropy energy density, obtained by an in-plane field sweep measurement using a vibrating sample magnetometer-superconducting quantum interference device (VSM-SQUID), as a function of the irradiation fluence, which is shown in Fig. \ref{fig:statics}e. The measurement result for such a field sweep for the as-deposited sample and the sample with irradiation fluence of  5$\times$10$^{15}$ ions/cm$^2$ are shown in the inset of  Fig. \ref{fig:statics}e.  We found that the PMA decreases monotonically with irradiation fluence, which is consistent with earlier studies \cite{Zhao:2019aa,Jong:2022ts}. Here, we specifically note that PMA persists up to a very large degree of intermixing (irradiation fluence), which is much higher than previous studies on ferromagnets\cite{Diez:2019aa,Zhao:2019aa,Juge:2021vd}. We attribute this to the lowered magnetic moment (demagnetization field) upon intermixing between Co and Gd, where more antiparallel (to Co) magnetic moment in Gd is induced.

Next, we both theoretically and experimentally investigated the effects of intermixing on AOS. We used a layered microscopic 3-temperature model (M3TM) \cite{Schellekens:2013aa} with angular momentum exchange between Co and Gd mediated by exchange scattering, which successfully described the AOS behaviour of both Co/Gd bilayers and CoGd alloys \cite{Beens:2019aa}. In our simulation, we keep the material parameters identical to those of our earlier works \cite{Beens:2019aa, Beens:2019ab}. In order to obtain the intermixing profile as a function of the He ion irradiation fluence, we carried out numerical simulations using TRIDYN~\cite{Moeller:1988}, which dynamically models the effects of ion irradiation on multilayer systems. 
We took into account the full layer stack and substrate, which was formerly adopted in earlier studies\cite{Jakubowski:2019, Krupinski:2021} successfully addressing  intermixing. Here, we specifically incorporated the element profile of Pt, Co and Gd (see Fig. \ref{fig:AOS}a) extracted from the TRIDYN simulations into our theoretical M3TM framework, where the Pt atoms are modeled as a magnetically dead component with the same exchange scattering properties as Gd. We found that the presence of Pt does neither significantly alter the AOS-ability nor the switching dynamics, which is governed by the Co/Gd interface. 
We further found that AOS still persists as long as the intermixing induced material gradient is present, moreover, as the pristine interface is gradually replaced by a CoGdPt alloy following the spatial composition gradient created by induced intermixing, 
the threshold fluence monotonically decreases (see Fig. \ref{fig:AOS}. b). 
To determine the physical origin of this change, we traced the reduction of the Curie temperature\cite{Devolder:2001vw,Beens:2019aa,Beens:2019ab} upon intermixing, which was found to be almost linearly decreasing with the threshold fluence up to about 4$\%$ for the irradiation fluence of 5$\times 10^{15}$ ions/cm$^2$. By artificially lowering the Curie temperature of the non-irradiated and lightly irradiated configuration by 4\%, we found a reduction of AOS switching energy only by $<1\times$10$^{8}$ J/m$^3$, which could not explain the more drastic reduction as shown in Fig. \ref{fig:AOS}b. Therefore, we attribute the reduction of the switching energy mainly to the change of switching dynamics incurred by the intermixing induced Co/Gd concentration gradient at the interface. As the degree of intermixing increases, effectively more efficient angular momentum transfer between Co and Gd takes place. On the other hand, the composition gradient ensures the presence of AOS by maintaining the typical switching mechanisms of a synthetic ferrimagnet\cite{Beens:2019aa,Gerlach:2017aa}.  

In order to experimentally verify the reduction of the AOS threshold fluence by He irradiation, we characterized the threshold fluence of irradiated samples of Ta(4)/Pt(4)/Co(1)/Gd(3)/TaN(4). Here, we adopted the approach used in Ref. \cite{Lalieu:2017aa}. We illuminated the sample with a single $\sim100\,$fs-laser pulse with various pulse energies and imaged the reversed domain using polar-Kerr microscopy. Some exemplary images are shown in the inset of Fig. \ref{fig:AOS}c, from which the threshold fluence is obtained by fitting the switched area as a function of the laser pulse energy. The obtained results are plotted in Fig. \ref{fig:AOS}c. We found that the threshold fluence of AOS reduces monotonically at first with respect to the He irradiation fluence, which is followed by a saturation at higher irradiation fluences. We found as much as 30$\%$ of reduction in the AOS threshold fluence for irradiation fluence below 2.5$\times$10$^{15}$ ions/cm$^2$, the qualitative trend of which matches our theoretical expectations. 
To unravel the origin of the saturation of AOS threshold fluence (with respect to the irradiation fluence) is beyond the scope of this work, but as the magnetic anisotropy for this fluence is low, this fluence region is less technologically relevant. 

In summary, we have explored the use of He ion irradiation to enhance AOS through the intermixing between the Co and Gd. 
We first presented the physical evidence of this intermixing as well as its effect on static magnetic properties. We show the square hysteresis is maintained upon He ion irradiation, while the anisotropy reduces monotonically. We theoretically predicted the reduction of AOS threshold fluence upon intermixing. We further experimentally verified that He ion irradiation leads to a 30$\%$ reduction in threshold fluence paving the way for local and high-throughput control of AOS at wafer-scale. 


\begin{acknowledgments}
This project has received funding from the European Union’s Horizon 2020 research and innovation programme under the Marie Skłodowska-Curie grant agreement No.860060 and No.861300. This project is also part of the research programme financed by the Netherlands Organisation for Scientific Research (NWO). 
\end{acknowledgments}
\bibliography{aipsamp}
\end{document}